\documentclass[aps,prl,reprint,groupedaddress]{revtex4-1}
\usepackage{amssymb}
\usepackage{amsmath}
\usepackage{array}
\usepackage{multirow,bigdelim}
\usepackage{enumitem}

\begin{document}

\title{n qubits can be entangled in two different ways}
\author{Dafa Li}

\begin{abstract}
Abstract: \ In [M. Walter et al., Science 340, 1205, 7 June (2013)], via
polytopes they gave a sufficient condition for genuinely entangled pure
states and discussed SLOCC classification. In this paper, we study
entanglement classification of pure states of n qubits via the basis state
matrix (BSM) whose rows are the basis states. We propose a canonical form of
BSM obtained by exchanging columns (i.e. permutation of qubits) and rows of
BSM and then a necessary and sufficient condition for a genuinely entangled
state of n qubits via a canonical form of BSM. Thus, for any n qubits, the
genuinely entangled states can be partitioned into two families. One family
includes all states whose BSM cannot be transformed into the canonical form.
The states with the BSM are always genuinely entangled no matter what the
non-zero coefficients are. GHZ and W states belong to the family. The other
includes all states whose BSM can be transformed into the canonical form,
but for any canonical form of BSM, some two columns or rows of the
corresponding coefficient matrix are not proportional. The cluster state
belongs to the family.

\end{abstract}



\affiliation{Department of Mathematical Sciences, Tsinghua University,
Beijing, 100084, China\\email: lidafa@tsinghua.edu.cn}

\maketitle

\section{Introduction}

Quantum entanglement is a key physical resource in quantum information and
computation such as quantum secure communication, entanglement swapping,
quantum cryptography, quantum teleportation, quantum dense coding, quantum
error correction and so on. A quantum state of $n$ qubits\ is genuinely
entangled\ if it cannot be written as $|\varphi \rangle |\phi \rangle $.
Detecting if a state is genuinely entangled\ is a challenging task in
quantum information theory and experiments \cite{Horodecki}. Many efforts
have contributed to understanding the different ways of entanglement \cite%
{Nielsen}.

It is known that it is hard to classify multipartite entanglement for four
or more qubits. D\"{u}r \emph{et al.} proposed that three qubits can be
entangled in two inequivalent ways: GHZ and W \cite{Dur}. In \cite%
{Verstraete}, Verstraete \emph{et al.} partitioned pure states of four
qubits into nine families. After then, SLOCC\ classification of four qubits
were studied deeply \cite{Miyake, Luque, Chterental, LDFQIC09, Ribeiro,
Buniy}.

For SLOCC classification of n qubits, the previous articles proposed
different SLOCC\ invariants: for example, the concurrence and 3-tangle \cite%
{Coffman}; polynomial\ invariants \cite{Miyake, Luque, Leifer, Levay,
LDF07a, Osterloh06, LDFJPA13}; the diversity degree and the degeneracy
configuration of a symmetric state \cite{Bastin}; ranks of coefficient
matrices \cite{LDFPRL12,LDFPRA12, FanJPA}; the entanglement polytopes \cite%
{Walter}.

Recently, spectra, standard Jordan normal form, and integer partitions\ were
used for SLOCC\ classification of $4n$ qubits \cite{LDFQIP18}. In \cite{SM
Fei, Jin}, they studied the genuine multipartite entanglement of arbitrary
n-partite quantum states.

In this paper, we propose a necessary and sufficient condition for a
genuinely entangled state of n qubits via a canonical form of BSM. Then, for
any n qubits, we partition the genuinely entangled states into two families.

\section{The basis state matrix (BSM)}

Let $|\psi \rangle $ be a pure state of $n$ qubits\ and $m$\ be the number
of non-zero coefficients of $|\psi \rangle $. Let $d_{i}\neq 0$. Then $|\psi
\rangle $ can be written as
\begin{equation}
|\psi \rangle =d_{1}|\nu _{1}^{(1)}\cdots \upsilon _{n}^{(1)}\rangle +\cdots
+d_{m}|\nu _{1}^{(m)}\cdots \upsilon _{n}^{(m)}\rangle .  \label{state}
\end{equation}%
where $\nu _{i}^{(s)}=0,1$. When $|\psi \rangle $ $=|r\rangle _{i}|\omega
\rangle _{12\cdots n/i}$, $r_{i}=0$ or $1$, for example $m=1$, clearly it is
separable. It is trivial to detect the state.

Let $B$ be the BSM of $|\psi \rangle $. Then,
\begin{equation}
B=\left(
\begin{array}{cccc}
\nu _{1}^{(1)} & \nu _{2}^{(1)} & \cdots & \upsilon _{n}^{(1)} \\
\cdots & \cdots & \cdots & \cdots \\
\nu _{1}^{(m)} & \nu _{2}^{(m)} & \cdots & \upsilon _{n}^{(m)}%
\end{array}%
\right) .
\end{equation}

Clearly, each row of the BSM $B$ corresponds to a basis state of $|\psi
\rangle $. The first column $\left(
\begin{array}{ccc}
\nu _{1}^{(1)} & \cdots & \nu _{1}^{(m)}%
\end{array}%
\right) ^{T}$ of $B$ corresponds to qubit 1 of these basis states.
Similarly, the ith column of $B$ corresponds to qubit $i$ of these basis
states. Therefore, exchanging the ith and the jth columns of BSM $B$ means
exchanging the ith and jth qubits. In other words, exchanging columns means
permutation of qubits. One can see that exchanging two rows of the BSM $B$
means exchanging two basis states.

Assume that $|\psi \rangle $ $\neq |r\rangle _{i}|\omega \rangle _{12\cdots
n/i}$, $r_{i}=0$ or $1$. By moving the columns $p_{1},p_{2},...,p_{k}$ of
the BSM $B$ to the left side of the BSM $B$ and then exchanging rows if
applicable, if $B$ is transformed into the following block matrix, then we
say that the block matrix in Eq. (\ref{blk-1}) is a canonical form of BSM $B$%
.
\begin{equation}
\left(
\begin{array}{cc}
\Pi _{1} & \Delta \\
\Pi _{2} & \Delta \\
\vdots & \vdots \\
\Pi _{g} & \Delta%
\end{array}%
\right) ,  \label{blk-1}
\end{equation}%
where
\begin{equation}
\Delta =\left(
\begin{array}{ccc}
j_{1}^{(1)} & \cdots & j_{l}^{(1)} \\
\cdots & \cdots & \cdots \\
j_{1}^{(h)} & \cdots & j_{l}^{h)}%
\end{array}%
\right) ,\Pi _{s}=\left(
\begin{array}{ccc}
i_{1}^{(s)} & \cdots & i_{k}^{(s)} \\
\cdots & \cdots & \cdots \\
i_{1}^{(s)} & \cdots & i_{k}^{(s)}%
\end{array}%
\right) ,  \label{blk-2}
\end{equation}%
where $s=1,2,...,$ $g$, $m=gh$, $\Pi _{i}\neq \Pi _{j}$ if $i\neq j$, and $%
g>1$ and $h>1$ because $|\psi \rangle $ $\neq |r\rangle _{i}|\omega \rangle
_{12\cdots n/i}$, $r_{i}=0$ or $1$.

For example, let $|\psi \rangle =(1/2)(|000\rangle +|010\rangle +|101\rangle
+|111\rangle )$. By exchanging the columns 2 \ and 1 (i.e. permutation of
qubits 1 and 2) of BSM $B$, then, exchanging rows 2 and 3, obtain the
following canonical form of $B$.

\begin{equation}
\left(
\begin{array}{cc}
\Pi _{1} & \Delta \\
\Pi _{2} & \Delta%
\end{array}%
\right) ,\text{ }\Delta =\left(
\begin{array}{cc}
0 & 0 \\
1 & 1%
\end{array}%
\right) ,\Pi _{1}=\left(
\begin{array}{c}
0 \\
0%
\end{array}%
\right) ,\Pi _{2}=\left(
\begin{array}{c}
1 \\
1%
\end{array}%
\right) .  \label{ex-1}
\end{equation}

We next demonstrate that $|\psi \rangle $ is separable via the canonical
form in Eq. (\ref{ex-1}). From $\left(
\begin{array}{cc}
\Pi _{1} & \Pi _{2}%
\end{array}%
\right) ^{T}$ and $\Delta $, we define $|\omega \rangle _{2}=\alpha
|0\rangle _{2}+\beta |1\rangle _{2})$\ and $|\varphi \rangle _{13}=\gamma
|00\rangle _{13}+\delta |11\rangle _{13}$, \ respectively. Clearly, when $%
\alpha =\beta =\gamma =\delta =1/\sqrt{2}$ , $|\omega \rangle _{2}|\varphi
\rangle _{13}=|\psi \rangle $.

\section{A necessary and sufficient condition for genuinely entangled states
of $n\geq 2$ qubits}

\textit{Theorem 1}. Let $|\psi \rangle $ be a pure state of $n$ qubits\ with
$m$ $(>1)$ non-zero coefficients in Eq. (\ref{state}). But, $|\psi \rangle
\neq |r\rangle _{i}|\omega \rangle _{12\cdots n/i}$, $r_{i}=0$ or $1$. $%
|\psi \rangle $ is separable if and only if BSM of $|\psi \rangle $ can be
transformed into the canonical form in Eq. (\ref{blk-1}) by moving some
columns of BSM to the left side of BSM and then exchanging rows if
applicable and any two rows or columns of the corresponding non-zero
coefficient matrix in Eq. (\ref{ma-eq-0}) are proportional.

Proof for the necessity..Assume that $|\psi \rangle $ is separable. Then,
let $|\psi \rangle $\ $=|\omega \rangle _{p_{1}\cdots p_{k}}|\varphi \rangle
_{q_{1}\cdots q_{l}}$, where $|\omega \rangle _{p_{1}\cdots p_{k}}$ and $%
|\varphi \rangle _{q_{1}\cdots q_{l}}$ are defined as follows.

\begin{eqnarray}
|\omega \rangle _{p_{1}\cdots p_{k}} &=&(\alpha _{1}|i_{1}^{(1)}\cdots
i_{k}^{(1)}\rangle +\cdots +\alpha _{g}|i_{1}^{(g)}\cdots i_{k}^{(g)}\rangle
)_{p_{1}\cdots p_{k}},  \label{sep-1} \\
|\varphi \rangle _{q_{1}\cdots q_{l}} &=&(\beta _{1}|j_{1}^{(1)}\cdots
j_{l}^{(1)}\rangle +\cdots +\beta _{h}|j_{1}^{(h)}\cdots j_{l}^{h)}\rangle
)_{q_{1}\cdots q_{l}},  \label{sep-2}
\end{eqnarray}

Clearly, $g>1$ and $h>1$ because $|\psi \rangle \neq |r\rangle _{i}|\omega
\rangle _{12\cdots n/i}$, $r_{i}=0$ or $1$. A calculation yields that

\begin{eqnarray}
&&|\omega \rangle _{p_{1}\cdots p_{k}}|\varphi \rangle _{q_{1}\cdots q_{l}}
\label{prod-1} \\
&=&\alpha _{1}\beta _{1}|\sigma _{1}^{(1)}\cdots \sigma _{n}^{(1)}\rangle
_{1\cdots n}+\cdots +\alpha _{1}\beta _{h}|\sigma _{1}^{(h)}\cdots \sigma
_{n}^{(h)}\rangle _{1\cdots n}  \notag \\
&&+\cdots +\alpha _{g}\beta _{1}|\sigma _{1}^{(g-1)h+1}\cdots \sigma
_{n}^{(g-1)h+1}\rangle _{1\cdots n}+\cdots \\
&&+\alpha _{g}\beta _{h}|\sigma _{1}^{(m)}\cdots \sigma _{n}^{(m)}\rangle
_{1\cdots n}  \notag \\
&=&|\psi \rangle .
\end{eqnarray}

Clearly, $gh=m$. Let us fill a $g$ by $h$ matrix with $m$ non-zero
coefficients of $|\psi \rangle $ as entries via Eqs. (\ref{state}, \ref%
{prod-1}). If $|\sigma _{1}^{(1)}\cdots \sigma _{n}^{(1)}\rangle =|\nu
_{1}^{(i)}\cdots \upsilon _{n}^{(i)}\rangle $, then $\alpha _{1}\beta
_{1}=d_{i}$ and let $a_{1}=d_{i}$. If $|\sigma _{1}^{(2)}\cdots \sigma
_{n}^{(2)}\rangle =|\nu _{1}^{(j)}\cdots \upsilon _{n}^{(j)}\rangle $, then $%
\alpha _{1}\beta _{2}=d_{j}$ and let $a_{2}=d_{j}$ and so on. Let $\alpha
=\left(
\begin{array}{cccc}
\alpha _{1} & \alpha _{2} & \cdots & \alpha _{g}%
\end{array}%
\right) $ and $\beta =\left(
\begin{array}{cccc}
\beta _{1} & \beta _{2} & \cdots & \beta _{h}%
\end{array}%
\right) $. Then, via Eqs. (\ref{state}, \ref{prod-1}), obtain the following
matrix equation.

\begin{equation}
\alpha ^{T}\beta =\left(
\begin{array}{cccc}
a_{1} & a_{2} & \cdots & a_{h} \\
a_{h+1} & a_{h+2} & \cdots & a_{2h} \\
\cdots & \cdots & \cdots & \cdots \\
a_{(g-1)h+1} & a_{(g-1)h+2} & \cdots & a_{m=gh}%
\end{array}%
\right) .  \label{ma-eq-0}
\end{equation}

It means that Eq. (\ref{ma-eq-0})\ has a solution for $\alpha _{i}$ and $%
\beta _{j}$. By Theorem 5 in Appendix A, any two columns or rows of the
matrix in Eq. (\ref{ma-eq-0}) are proportional.

We next study BSM obtained from Eq. (\ref{prod-1}). By moving the columns $%
p_{1},p_{2},...,p_{k}$ of BSM to the left side of BSM and then exchanging
rows if applicable, obtain the following matrix
\begin{equation}
\left(
\begin{array}{cccccc}
i_{1}^{(1)} & \cdots & i_{k}^{(1)} & j_{1}^{(1)} & \cdots & j_{l}^{(1)} \\
\cdots & \cdots & \cdots & \cdots & \cdots & \cdots \\
i_{1}^{(1)} & \cdots & i_{k}^{(1)} & j_{1}^{(h)} & \cdots & j_{l}^{h)} \\
\cdots & \cdots & \cdots & \cdots & \cdots & \cdots \\
i_{1}^{(g)} & \cdots & i_{k}^{(g)} & j_{1}^{(1)} & \cdots & j_{l}^{(1)} \\
\cdots & \cdots & \cdots & \cdots & \cdots & \cdots \\
i_{1}^{(g)} & \cdots & i_{k}^{(g)} & j_{1}^{(h)} & \cdots & j_{l}^{h)}%
\end{array}%
\right) .
\end{equation}

The above matrix can be written as the block matrix. Thus, we obtain the
canonical form.

Proof for the sufficiency. Assume that BSM of $|\psi \rangle $ can be
transformed into the canonical form in Eq. (\ref{blk-1}) by moving the
columns $p_{1}$, $\cdots ,p_{k}$ of BSM of $|\psi \rangle $ to the left side
of BSM and exchanging rows if applicable. Note that $\left(
\begin{array}{ccc}
\Pi _{1} & \cdots & \Pi _{g}%
\end{array}%
\right) ^{T}$ corresponds to qubits $p_{1}$, $\cdots ,p_{k}$ and $\left(
\begin{array}{ccc}
\Delta & \cdots & \Delta%
\end{array}%
\right) ^{T}$ corresponds to qubits $q_{1},\cdots ,q_{l}$. Via $\left(
\begin{array}{ccc}
\Pi _{1} & \cdots & \Pi _{g}%
\end{array}%
\right) ^{T}$ and $\Delta $, we define $|\omega \rangle _{p_{1}\cdots p_{k}}$
and $|\varphi \rangle _{q_{1}\cdots q_{l}}$ in Eqs. (\ref{sep-1}, \ref{sep-2}%
), \ respectively.

To show that $|\psi \rangle $ is separable, we need to check if there are $%
\alpha _{i}$ and $\beta _{j}$ such that $|\omega \rangle _{p_{1}\cdots
p_{k}}|\varphi \rangle _{q_{1}\cdots q_{l}}=|\psi \rangle $. To this end, a
calculation yields Eq. (\ref{prod-1}). Thus, via Eq. (\ref{prod-1})\ obtain
Eq. (\ref{ma-eq-0}). By Theorem 5 in Appendix A, Eq. (\ref{ma-eq-0}) has a
solution for $\alpha _{i}$ and $\beta _{j}$. Therefore, $|\omega \rangle
_{p_{1}\cdots p_{k}}|\varphi \rangle _{q_{1}\cdots q_{l}}=|\psi \rangle $.
That is, $|\psi \rangle $ is separable. Q.E.D.

For example. Let $|\varpi \rangle =\sum_{i=0}^{2^{n}-1}c_{i}|i\rangle $,
where $c_{i}\neq 0$, be a state of $n$ qubits. By Theorem 1, $|\varpi
\rangle $ is separable if and only if $\frac{c_{0}}{c_{2^{n-1}}}=\frac{c_{1}%
}{c_{2^{n-1}+1}}=\cdots =\frac{c_{2^{n-1}-1}}{c_{2^{n}-1}}$.

Restated in the contrapositive Theorem 1 reads

\textit{Theorem 2.} Let $|\psi \rangle $ be a state in Eq. (\ref{state}).
But, $|\psi \rangle \neq |r\rangle _{i}|\omega \rangle _{12\cdots n/i}$, $%
r_{i}=0$ or $1$. $|\psi \rangle $ is genuinely entangled if and only if BSM
cannot be transformed into the canonical form in Eq. (\ref{blk-1}) by moving
some columns of BSM to the left side of BSM and then exchanging rows if
applicable or for any canonical form of BSM the corresponding non-zero
coefficient matrix in Eq. (\ref{ma-eq-0}) always has some two rows or
columns which are not proportional.

By Theorem 2, GHZ, W, Dicke, GHZ+Dicke, and cluster states of $n$ qubits, $%
|\Phi _{2}\rangle $, $\ |\Phi _{4}\rangle $,$\ \ |\Phi _{5}\rangle $, $|\Psi
_{2}\rangle $, $|\Psi _{4}\rangle $, $|\Psi _{5}\rangle $, $|\Psi
_{6}\rangle $, $|\Xi _{2}\rangle $, $|\Xi _{4}\rangle $, $|\Xi _{5}\rangle $%
, $|\Xi _{6}\rangle $, $|\Xi _{7}\rangle $ are genuinely entangled \cite%
{Osterloh06}.

\textit{Corollary} 1. Let $|\psi \rangle $ be a pure state with $m$\ ($>1$)
non-zero coefficients\ of $n$ qubits. Assume that $|\psi \rangle \neq
|r\rangle _{i}|\omega \rangle _{12\cdots n/i}$, $r_{i}=0$ or $1$. If $m$ is
a prime, then $|\psi \rangle $ is genuinely entangled .

By Corollary 1, $|\Phi _{2}\rangle $, $|\Phi _{5}\rangle $, $|\Psi
_{2}\rangle $, $|\Psi _{5}\rangle $, $|\Xi _{2}\rangle $, $|\Xi _{5}\rangle $%
, $|\Xi _{7}\rangle $ are genuinely entangled \cite{Osterloh06}.

\section{ A necessary and sufficient condition for genuinely entangled
states of $n\geq 2$ qubits and $m=4$}

It is known that $0$ and $1$ are complementary. Let $1-\gamma =\gamma
^{\prime }$, where $\gamma =0,1$. Then, we say that $\gamma ^{\prime }$ and $%
\gamma $\ are complementary. We call $|i_{1}i_{2}\cdots i_{n}\rangle $\ \
and $|i_{1}^{\prime }i_{2}^{\prime }\cdots i_{n}^{\prime }\rangle $, for
example $|0101\rangle $ and $|1010\rangle $, a pair of complementary basis
states.

\textit{Theorem 3}. Let $|\psi \rangle $ be a pure state\ of $n$ qubits and $%
m=4$. Assume that $|\psi \rangle \neq |r\rangle _{i}|\omega \rangle
_{12\cdots n/i}$, $r_{i}=0$ or $1$. Let $d_{i}\neq 0$. $|\psi \rangle $ can
be written as
\begin{eqnarray}
|\psi \rangle &=&d_{1}|\nu _{1}^{(1)}\cdots \upsilon _{n}^{(1)}\rangle
+d_{2}|\nu _{1}^{(2)}\cdots \upsilon _{n}^{(2)}\rangle  \notag \\
&&+d_{3}|\nu _{1}^{(3)}\cdots \upsilon _{n}^{(3)}\rangle +d_{4}|\nu
_{1}^{(4)}\cdots \upsilon _{n}^{(4)}\rangle ,  \label{state-m-4}
\end{eqnarray}%
\ where $\nu _{1}^{(1)}\cdots \upsilon _{n}^{(1)}<\nu _{1}^{(2)}\cdots
\upsilon _{n}^{(2)}<\nu _{1}^{(3)}\cdots \upsilon _{n}^{(3)}<\nu
_{1}^{(4)}\cdots \upsilon _{n}^{(4)}$. Then, $|\psi \rangle $\ is separable
if and only if $|\psi \rangle $ has two pairs of complementary basis states
and $d_{1}d_{4}=d_{2}d_{3}$.

Proof for the necessity. Assume that $|\psi \rangle $\ is separable. Then,
let
\begin{equation}
|\psi \rangle =(\alpha _{1}|\gamma _{1}\cdots \gamma _{k}\rangle +\alpha
_{2}|\delta _{1}\cdots \delta _{k}\rangle )(\beta _{1}|\iota _{1}\cdots
\iota _{s}\rangle +\beta _{2}|\varepsilon _{1}\cdots \varepsilon _{s}\rangle
),
\end{equation}%
where $\gamma _{i}$, $\delta _{i}$, $\iota _{i}$, $\varepsilon _{i}$ are 0
or 1. Clearly, $\gamma _{\ell }\neq \delta _{\ell }$, $\ell =1,2,...,k$,
because $|\psi \rangle \neq |r\rangle _{i}|\omega \rangle _{12\cdots n/i}$, $%
r_{i}=0$ or $1$. So, $\delta _{\ell }=1-\gamma _{\ell }=\gamma _{\ell
}^{\prime }$, $\ell =1,2,...,k$. Similarly, $\varepsilon _{\ell }=t_{\ell
}^{\prime }$, $\ell =1,2,\cdots ,s$. Then,
\begin{equation}
|\psi \rangle =(\alpha _{1}|\gamma _{1}\cdots \gamma _{k}\rangle +\alpha
_{2}|\gamma _{1}^{\prime }\cdots \gamma _{k}^{\prime }\rangle )(\beta
_{1}|t_{1}\cdots t_{s}\rangle +\beta _{2}|t_{1}^{\prime }\cdots
t_{s}^{\prime }\rangle ).  \notag
\end{equation}%
One can see that $|\psi \rangle $ \ has two pairs of complementary basis
states.

Eq. (\ref{ma-eq-0}) reduces to
\begin{equation}
\left(
\begin{array}{c}
\alpha _{1} \\
\alpha _{2}%
\end{array}%
\right) \left(
\begin{array}{cc}
\beta _{1} & \beta _{2}%
\end{array}%
\right) =\left(
\begin{array}{cc}
d_{1} & d_{2} \\
d_{3} & d_{4}%
\end{array}%
\right) \text{ or }\left(
\begin{array}{cc}
d_{1} & d_{3} \\
d_{2} & d_{4}%
\end{array}%
\right)
\end{equation}

By Theorem 1 and from Theorem 5 in Appendix A, $d_{1}d_{4}=d_{2}d_{3}$.

Proof for the sufficiency.\ Assume that $|\psi \rangle $ has two pairs of
complementary basis states and $d_{1}d_{4}=d_{2}d_{3}$. Let
\begin{equation}
|\psi \rangle =d_{1}|i_{1}i_{2}\cdots i_{n}\rangle +d_{2}|j_{1}\cdots
j_{n}\rangle +d_{3}|j_{1}^{\prime }\cdots j_{n}^{\prime }\rangle
+d_{4}|i_{1}^{\prime }\cdots i_{n}^{\prime }\rangle .  \label{comp-1}
\end{equation}

Let $|i_{1}\cdots i_{n}\rangle =|l_{q_{1}}\cdots l_{q_{t}}\rangle
|l_{p_{1}}\cdots l_{p_{s}}\rangle $ and $|j_{1}\cdots j_{n}\rangle
=|l_{q_{1}}\cdots l_{q_{t}}\rangle |l_{p_{1}}^{\prime }\cdots
l_{p_{s}}^{\prime }\rangle $. Then $|j_{1}^{\prime }\cdots j_{n}^{\prime
}\rangle =|l_{q_{1}}^{\prime }\cdots l_{q_{t}}^{\prime }\rangle
|l_{p_{1}}\cdots l_{p_{s}}\rangle $, and $|i_{1}^{\prime }\cdots
i_{n}^{\prime }\rangle =|l_{q_{1}}^{\prime }\cdots l_{q_{t}}^{\prime
}\rangle |l_{p_{1}}^{\prime }\cdots l_{p_{s}}^{\prime }\rangle $. Let $\frac{%
d_{1}}{d_{2}}=\frac{d_{3}}{d_{4}}=\mu $. Then, $d_{1}=\mu d_{2}$ and $%
d_{3}=\mu d_{4}$. Then, via Eq. (\ref{comp-1}), $|\psi \rangle $ can be
rewritten as

\begin{equation}
|\psi \rangle =(d_{2}|l_{q_{1}}\cdots l_{q_{t}}\rangle
+d_{4}|l_{q_{1}}^{\prime }\cdots l_{q_{t}}^{\prime }\rangle )(\mu
|l_{p_{1}}\cdots l_{p_{s}}\rangle +|l_{p_{1}}^{\prime }\cdots
l_{p_{s}}^{\prime }\rangle ).
\end{equation}

Thus, $|\psi \rangle $ is separable. Q.E.D.

Restated in the contrapositive Theorem 3 reads

\textit{Theorem 4}. Let $|\psi \rangle $ be a pure state\ of $n$ qubits in
Eq. (\ref{state-m-4}). Assume that $|\psi \rangle \neq |r\rangle _{i}|\omega
\rangle _{12\cdots n/i}$, $r_{i}=0$ or $1$. $|\psi \rangle $ is genuinely
entangled if and only if $|\psi \rangle $ does not have two pairs of
complementary basis states, or $d_{1}d_{4}\neq d_{2}d_{3}$.

By Theorem 4, it is trivial to know that $|\Phi _{4}\rangle $, $|\Psi
_{4}\rangle $, $|\Xi _{4}\rangle $ are genuinely entangled\ \cite{Osterloh06}%
.

\section{Entanglement classification}

We partition pure states of $n$ qubits into the following four families.

Family (1) (separable). It includes $|\psi \rangle =|r\rangle _{i}|\omega
\rangle _{12\cdots n/i}$, $r_{i}=0$ or $1$.

Family (2) (separable). It includes $|\psi \rangle $ ( $|\psi \rangle \neq
|r\rangle _{i}|\omega \rangle _{12\cdots n/i}$, $r_{i}=0$ or $1$) whose BSM
becomes a canonical form and any two rows or columns of the corresponding
non-zero coefficient matrix\ are proportional

Family (3) (genuinely entangled based on only BSM). It includes the state $%
|\psi \rangle $ ($|\psi \rangle \neq |r\rangle _{i}|\omega \rangle
_{12\cdots n/i}$, $r_{i}=0$ or $1$) whose BSM cannot become canonical form
by moving some columns of BSM to the left side of BSM and then exchanging
rows if applicable.

The states whose BSM cannot become canonical form are always genuinely
entangled no matter what the non-zero coefficients are. It means that the
entanglement is independent to the coefficients. For example, $\alpha
|0\cdots 0\rangle +\beta |1\cdots 1\rangle $ is always genuinely entangled
for any non-zero $\alpha $ and $\beta $. GHZ, W, and Dicke states of $n$
qubits belong to Family 3. The state $|\psi \rangle $ ($|\psi \rangle \neq
|r\rangle _{i}|\omega \rangle _{12\cdots n/i}$, $r_{i}=0$ or $1$) with prime
$m$ also belongs to Family 3.

Family (4) (genuinely entangled based on BSM and coefficients). It includes
the state $|\psi \rangle $ ($|\psi \rangle \neq |r\rangle _{i}|\omega
\rangle _{12\cdots n/i}$, $r_{i}=0$ or $1$) whose BSM can be transformed
into the canonical form by moving some columns of BSM to the left side of
BSM and then exchanging rows if applicable, but for any canonical form of
BSM, some two columns or rows of the corresponding coefficient matrix are
not proportional.

The cluster state belongs to Family 4. For the cluster state $|C_{4}\rangle =%
\frac{1}{2}|0000\rangle +\frac{1}{2}|0101\rangle +\frac{1}{2}|1010\rangle -%
\frac{1}{2}|1111\rangle $, if $-1/2$ is changed as $1/2$, then it becomes
separable. The entanglement for the family 4 is related to the coefficients.
The states of Family 4 are referred as to general cluster states.

\textit{Remark 1}. For two qubits, Bell state belongs to Family 3 and $C_{2}=%
\frac{1}{2}|00\rangle +\frac{1}{2}|01\rangle +\frac{1}{2}|10\rangle -\frac{1%
}{2}|11\rangle $ belongs to Family 4. It is known that Bell state and $C_{2}$
are equivalent under SLOCC. So, the entanglement classification in this
paper is not SLOCC classification.

\section{Summary}

In this paper, we study entanglement classification of pure states of n
qubits via BSM and coefficient matrices. We propose a canonical form of BSM
obtained by exchanging columns (i.e. permutation of qubits) and rows. We
propose a necessary and sufficient condition for a genuinely entangled
state. Thus, for any n qubits, the genuinely entangled states can be
partitioned into two families. One family includes all states whose BSM
cannot be transformed into the canonical form by exchanging columns (i.e.
permutation of qubits) and rows. The other includes all states whose BSM can
be transformed into the canonical form, but for any canonical form of BSM,
some two columns or rows of the corresponding coefficient matrix are not
proportional.

\section{Statements and declarations}

Non - financial interests, no competng interests, no finacila supports.

\section{Appendix A. A matrix equation}

Let $X=\left(
\begin{array}{cccc}
x_{1} & x_{2} & \cdots & x_{k}%
\end{array}%
\right) $ and $Y=\left(
\begin{array}{cccc}
y_{1} & y_{2} & \cdots & y_{\ell }%
\end{array}%
\right) $. Let us solve the following matrix equation.

\begin{equation}
X^{T}Y=\left(
\begin{array}{cccc}
a_{1} & a_{2} & \cdots & a_{\ell } \\
a_{\ell +1} & a_{\ell +2} & \cdots & a_{2\ell } \\
\cdots & \cdots & \cdots & \cdots \\
a_{(k-1)\ell +1} & a_{(k-1)\ell +2} & \cdots & a_{k\ell }%
\end{array}%
\right) =A,  \label{mat-eq-3}
\end{equation}%
where $a_{i}\neq 0$ for any $i$.

\textit{Theorem 5 }(version 1).. Eq. (\ref{mat-eq-3}) has a solution if and
only if any two rows and columns of $A$\ are proportional, respectively.

Proof. Assume that Eq. (\ref{mat-eq-3}) has a solution. Then, from the first
and second columns of Eq. ( \ref{mat-eq-3}), we obtain $\frac{y_{1}}{y_{2}}=%
\frac{a_{1}}{a_{2}}=\frac{a_{\ell +1}}{a_{\ell +2}}=\cdots =\frac{%
a_{(k-1)\ell +1}}{a_{(k-1)\ell +2}}$. Similarly, we can show any two columns
of Eq. (\ref{mat-eq-3}) are proportional. Similarly, we can show that any
two rows of Eq. (\ref{mat-eq-3}) are proportional.

Conversely, assume that any two rows and any two columns of Eq. (\ref%
{mat-eq-3}) are proportional, respectively. Then, we show that Eq. (\ref%
{mat-eq-3}) has a solution below. Let $\tau _{12}$ be the ratio of the first
two columns of Eq. (\ref{mat-eq-3}). Then, $\frac{y_{1}}{y_{2}}=\tau _{12}$.
Similarly, we define $\tau _{13}$ and $\tau _{1\ell }$. One can know that $%
\frac{y_{1}}{y_{3}}=\tau _{13}$ and $\frac{y_{1}}{y_{\ell }}=$ $\tau _{1\ell
}$. Next, let $\sigma _{12}$ be the ratio of the first two rows of Eq. (\ref%
{mat-eq-3}). Then, $\frac{x_{1}}{x_{2}}=\sigma _{12}$. Similarly, we define $%
\sigma _{13}$ and $\sigma _{1k}$ . Then, $\frac{x_{1}}{x_{3}}=\sigma _{13}$
and $\frac{x_{1}}{x_{k}}=\sigma _{1k}$. One can see that as soon as $x_{1}$ (%
$y_{1}$) is given, all $x_{j}$ ($y_{j}$) are known. Clearly, that $%
x_{1}y_{1}=$ $a_{1}$ has many solutions for $x_{1}$ and $y_{1}$. So, Eq. (%
\ref{mat-eq-3}) has many solutions. Q.E.D.

\textit{Remark 2}. One can show that any two rows of $A$ are proportional if
and only if any two columns are proportional.

So, we have the second version of Theorem 5 as follows.

\textit{Theorem 5 }(version 2). Eq. (\ref{mat-eq-3}) has a solution if and
only if any two rows or columns of $A$\ are proportional.


\begin{thebibliography}{99}
\bibitem{Horodecki} R. Horodecki et al., Rev. Mod. Phys. 81, 865 (2009).

\bibitem{Nielsen} M.A. Nielsen and I.L. Chuang, \textsl{Quantum Computation
and Quantum Information} (Cambridge Univ. Press, Cambridge, 2000).

\bibitem{Dur} W. D\"{u}r, G. Vidal, and J.I. Cirac, Phys. Rev. A \textbf{62}%
, 062314 (2000).

\bibitem{Verstraete} F. Verstraete, J. Dehaene, B. De Moor, and H.
Verschelde, Phys. Rev. A \textbf{65}, 052112 (2002).

\bibitem{Miyake} A. Miyake, Phys. Rev. A 67, 012108 (2003).

\bibitem{Luque} J.-G. Luque and J.-Y. Thibon, Phys. Rev. A \textbf{67},
042303 (2003).

\bibitem{Chterental} O. Chterental and D.Z. Djokovi\'{c}, in Linear Algebra
Research Advances, edited by G.D. Ling (Nova Science Publishers, Inc.,
Hauppauge, NY, 2007), Chap. \textbf{4}, 133.

\bibitem{LDFQIC09} D. Li, X. Li, H. Huang, and X. Li, Quantum Inf. Comput.
\textbf{9}, 0778 (2009).

\bibitem{Ribeiro} P. Ribeiro and R. Mosseri, Phys. Rev. Lett. \textbf{106},
180502 (2011).

\bibitem{Buniy} R.V. Buniy and T.W. Kephart, J. Phys. A: Math. Theor.
\textbf{45}, 185304 (2012).

\bibitem{Coffman} V. Coffman, J. Kundu, and W.K. Wootters, Phys. Rev. A
\textbf{61}, 052306 (2000).

\bibitem{Leifer} M.S. Leifer, N. Linden and A. Winter, Phys. Rev. A \textbf{%
69}, 052304 (2004).

\bibitem{Osterloh06} A. Osterloh and J. Siewert, international J. of quantum
information 4, 531 (2006).

\bibitem{Levay} P. Levay, J. Phys. A: Math. Gen. \textbf{39}, 9533, (2006).

\bibitem{LDF07a} D. Li, X. Li, H. Huang, and X. Li, Phys. Rev. A \textbf{76}
, 032304 (2007).

\bibitem{LDFJPA13} X. Li and D. Li, J. Phys. A: Math. Theor. \textbf{46},
135301, (2013).

\bibitem{Bastin} T. Bastin, S. Krins, P. Mathonet, M. Godefroid, L. Lamata,
and E. Solano, Phys. Rev. Lett. \textbf{103}, 070503 (2009).

\bibitem{LDFPRL12} X. Li and D. Li, Phys. Rev. Lett. \textbf{108}, 180502
(2012).

\bibitem{LDFPRA12} X. Li and D. Li, Phys. Rev. A \textbf{86}, 042332 (2012).

\bibitem{FanJPA} Bo Li, L.C. Kwek, and Heng Fan, J. Phys. A: Math. Theor.
\textbf{45}, 505301 (2012).\ \

\bibitem{Walter} M. Walter, B. Doran, D. Gross, and M. Christandl, Science
340, 1205, 7 June (2013).

\bibitem{LDFQIP18} Dafa Li, Quantum Inf Process 17, 1 (2018).

\bibitem{SM Fei} Yu Lu \ and Shao-Ming Fei, Results In Physics 52
(2023)106816

\bibitem{Jin} Zhi-Xiang Jin et al., Results in Physics 44 (2023) 106155.
\end{thebibliography}
\end{document}